\documentstyle[aps,preprint,epsf,epsfig,rotate,prl]{revtex}
\tightenlines
\newcommand{\be}{\begin{equation}}
\newcommand{\ee}{\end{equation}}
\newcommand{\ba}{\begin{eqnarray}}
\newcommand{\ea}{\end{eqnarray}}

\draft        
\begin{document}             
\title{Collective Modes in a Dilute Bose-Fermi Mixture}
\author{S. K. Yip}
\address{Institute of Physics, Academia Sinica, Nankang, Taipei 11529, Taiwan}
\maketitle

\begin{abstract}
We here study the collective excitations of a dilute
spin-polarized Bose-Fermi mixture
at zero temperature, considering in particular the features arising
from the interaction between the two species.
We show that
a propagating zero-sound mode is possible for the fermions
even when they do not interact among themselves.


\end{abstract}

Recent experimental progress in atomically trapped gases
has led to a resurgence of interest in quantum fluids.
A particular notable feature is the number of systems available,
ranging from single component bose gas in the original
experiments where Bose-Einstein Condensation (BEC) was first achieved
\cite{Anderson95} to binary bose mixture \cite{binary},
spinor condensate in optical traps \cite{Stenger98} and
degenerate fermi gas \cite{DeMarco99a,DeMarco99b}.  
Other systems have also received much recent attention,
in particular bose-fermi mixture.
This last mentioned system occurs naturally if `sympathetic cooling'
is employed to reduce the kinetic energy of the fermions
\cite{Mewes00}.  
There have already been several studies on the properties of this
system.  Questions addressed include stability against 
phase separation \cite{Viverit00,Bijlsma00} and collective excitations 
\cite{Bijlsma00}.

Although Bose-Fermi mixtures have been studied intensively
in low temperature physics in the context of $^3$He-$^4$He mixtures
\cite{Edwards92},
the atomically trapped gases offer many additional possibilities.
By the choice of atoms, concentration
of the various components, or the control of interaction
strength   among them by external fields \cite{Cornish00},
one can unmask phenomena previously unobservable.
In this paper, we shall study one example of this by
considering the density oscillations
of a bose-fermi mixture at low temperatures.  We shall
show that a variety of novel phenomena can arise
due to the coupling between the two components
for suitable parameters such as the ratio of
the sound velocity of the bose gas to the fermi velocity
of the fermions.
In particular, we shall show that it is possible to have
a propagating fermionic sound mode even in the absence
of  interaction among the fermions themselves.
Sound propagation has also been considered in ref \cite{Bijlsma00}
which however did not investigate the effects being studied here.
We shall comment on this later.

We shall then consider a mixture of weakly interacting
bose and fermi gases at zero temperature.  Both gases
are assumed to be spin-polarized such as would be
the case usually in magnetic traps.   For a dilute
mixture, interaction among the bosons themselves
and between the bosons and fermions can be characterised
by the scattering lengths $a_{bb}$ and $a_{bf}$ in
the s-wave channels.  The fermions however do not
interact among themselves since they are spin-polarized.
For simplicity we shall consider a uniform system.
We shall further assume that the gas is stable against
phase-separation unless explicitly specified.
We are interested in the density waves of this system.
As we shall see in general the modes may be damped.
Also since the density oscillations are likely to
be studied by exciting the systems with external potentials,
we shall instead consider the density responses of
the system under external perturbing potentials.
Collective modes of the system will show up as resonances
of these responses.

The Hamiltonian density is given by

\ba
H &=&   
 {\hbar^2 \over 2 m_b}  \nabla \psi_b^{\dagger} \ \nabla \psi_b  
  - \mu_b \psi_b^{\dagger} \psi_b + 
 {\hbar^2 \over 2 m_f}  \nabla \psi_f^{\dagger} \ \nabla \psi_f 
  - \mu_f \psi_f^{\dagger} \psi_f 
\nonumber \\
 & &  + { 1 \over 2} g_{bb} \psi_b^{\dagger} \psi_b^{\dagger} \psi_b \psi_b
  +  g_{bf} \psi_b^{\dagger} \psi_f^{\dagger} \psi_f \psi_b
  \nonumber  \\ 
  & &  \qquad \qquad +  \psi_b^{\dagger} \psi_b V_b^{\rm ext}
  +  \psi_f^{\dagger} \psi_f V_f^{\rm ext}
\label{H}
\ea
where the subscripts $b$ and $f$ denote bosons and fermions
respectively, $\psi_f$, $\psi_b$ are the field operators, 
$m_b$, $m_f$ the masses,
$\mu_b$, $\mu_f$ are the chemical potentials,
 $V_b^{\rm ext}$
and $V_f^{\rm ext}$ the external potentials.
All $\psi$'s and $V^{\rm ext}$'s
 are implicitly at the same physical point $\vec r$ in space.
The interaction parameters $g_{bb}$ and $g_{bf}$ are related to
the scattering lengths $a_{bb}$ and $a_{bf}$ by
$g_{bb} = 4 \pi \hbar^2 a_{bb}/ m_{bb}$ and
 $g_{bf} = 2 \pi \hbar^2 a_{bf} / m_r$
where $m_r$ is the reduced mass 
[ $m_r^{-1} \equiv  m_f^{-1} + m_b^{-1}   $].
 
We shall treat the interaction $g_{bb}$ and $g_{bf}$ within 
the Bogoliubov and random phase
approximation respectively \cite{Fetter}.
The results can be written in the physically transparent 
form:

\ba
\delta n_b (q, \omega) &=& - \chi_b [ g_{bf} \delta n_f + V^{\rm ext}_b ] \\
\nonumber
\delta n_f (q, \omega) &=& - \chi_f [ g_{bf} \delta n_b + V^{\rm ext}_f ] 
\label{r1}
\ea

\noindent expressing the response of the bosons and fermions
to the potentials due to the other species and the 
external perturbations (the terms in the square brackets).
Here $\delta n_b (q, \omega)$, $\delta n_f (q, \omega)$
are the deviations of the bosonic and fermionic densities
from equilibrium at wavevector $q$ and frequency $\omega$,

\be
\chi_b = - { 1 \over g_{bb} } \ 
             \left[ { c_b^2 q^2 \over \omega^2 - c_b^2 q^2 - 
                  (q^2/2 m_b)^2 } \right]
\label{chib}
\ee
\noindent and
\be
\chi_f = N_f \  \left[ 1 - {\omega \over 2 v_f q } 
   {\rm ln}
      ({\omega + v_f q \over \omega - v_f q}) \right]
\label{chif}
\ee

\noindent are the ($q$ and $\omega$ dependent) 
responses of the pure bosons and fermions
systems respectively to effective external potentials.  
$N_f \equiv p_f m_f / 2 \pi^2 $ is the density of
states for the fermions. 
($ p_f = (6 \pi^2 n_f)^{1/3}$ is the fermi momentum,
$v_f = p_f / m_f$)
For simplicity, in eq (\ref{chif}) I have already left out
terms that are small if $q << p_f$. 
$\omega$ should be interpreted as having a small and
positive part.

eq(\ref{r1}) can be re-arranged as
\begin{equation}
\left(
\begin{array}{cc}
1 &  g_{bf} \chi_b \\
g_{bf} \chi_f  &  1
\end{array}
\right)
\left(
\begin{array}{c}
\delta n_b \\
\delta n_f
\end{array}
\right)
\ = \ - 
\left(
\begin{array}{c}
\chi_b V_b^{\rm ext} \\
\chi_f V_f^{\rm ext}
\end{array}
\right)
\label{r2}
\end{equation}

Then finally

\begin{equation}
\left(
\begin{array}{c}
\delta n_b \\
\delta n_f
\end{array}
\right)
\ = \  - \ 
{1 \over 1 - g_{bf}^2 \chi_b \chi_f }
\left(
\begin{array}{cc}
1 &  - g_{bf} \chi_b \\
-g_{bf} \chi_f  &  1
\end{array}
\right)
\left(
\begin{array}{c}
\chi_b V_b^{\rm ext} \\
\chi_f V_f^{\rm ext}
\end{array}
\right)
\label{final}
\end{equation}

In the case where $g_{bf} = 0$, $\delta n_b = - \chi_b V^{\rm ext}_b$
and $\delta n_f = - \chi_f V^{\rm ext}_f$ and the responses
thus reduce to those of the pure bose and fermi gases.
The corresponding formulas for $\chi_b$ and $\chi_f$ were
already given in eq (\ref{chib}) and eq (\ref{chif}) above.
Before we proceed we shall recall the behavior of these
responses \cite{Fetter} and thus the collective modes.
For simplicity we shall restrict ourselves to small wavevectors,
{\it i.e.} $ q << m_b c_b$ and $ p_f$, and without
loss of generality $\omega > 0$.
The bosonic response ${\rm Im } \chi_b$ consists of a delta function at the
 excitation frequency $\omega =  c_b q$.
This is due to the Bogoliubov mode which
is purely propagating and undamped.
For the fermions however, there is no collective behavior.
The absorptive part, ${\rm Im} \chi_f$, is finite for
a whole range of frequencies $ | \omega | < v_f q$,
known as the particle-hole continuum, arising from the many possibilities
of independent particle-hole excitations.  ${\rm Re} \ \chi_b$ is simple.
It is given by $g_{bb}^{-1}$ at $\omega = 0$ and diverges
to $\pm \infty$ as $\omega \rightarrow c_b q $ from below and above
respectively.  ${\rm Re} \chi_f$ is given by $N_f$ at 
$\omega = 0$.  It decreases with increasing $\omega$, changes
sign at around $\omega \sim 0.83 v_f q$ and approaches $- \infty$
as $\omega \rightarrow v_f q$ from both above and below.
  For $\omega > v_f q$, it
remains negative with its magnitude gradually approaching zero
as $\omega \to \infty$.

Now we return to the bose-fermi mixture.
The response $\delta n_b$ to an external potential $V_b^{\rm ext}$
acting on the bosons only is
given by $ \chi_b / ( 1 - g_{bf}^2 \chi_b \chi_f )$.
The existence and the dispersion of the bosonic collective mode are
determined by
the solution to the equation 
$(\chi_b)^{-1} - g_{bf}^2 \chi_f = 0$, {\it i.e.}
\be
[ - \omega ^2 + c_b^2 q^2 + ({q^2 \over 2 m_b})^2 ]
- \left( {g_{bf}^2 \over g_{bb} }\right)
  ( c_b^2 q^2 ) \chi_f = 0
\label{bmode}
\ee

It will be convenient to discuss the normalized response
\be
\tilde \chi_b \equiv g_{bb}  \chi_b / ( 1 - g_{bf}^2 \chi_b \chi_f )
\label{nb}
\ee
\noindent $\tilde \chi_b = 1$ in the
static limit ($\omega = 0$, $q \rightarrow 0$) 
when there is no boson-fermion
interaction ($g_{bf} = 0$).

Similiarly the fermionic response to an external potential
acting on the fermions alone is 
$\chi_f / ( 1 - g_{bf}^2 \chi_b \chi_f)$.  
We shall discuss the behavior of the normalized quantity
\be
\tilde \chi_f \equiv N_f^{-1} \chi_f /  ( 1 - g_{bf}^2 \chi_b \chi_f )
\label{nf}
\ee
\noindent The normalization is chosen such 
that $\tilde \chi_f = 1$ in the
static limit ($\omega = 0$, $q \rightarrow 0$) 
when there is no boson-fermion
interaction ($g_{bf} = 0$).

Before proceeding let us first examine the responses at $\omega = 0$.
Stability requires that the density responses
$ \chi_b / ( 1 - g_{bf}^2 \chi_b \chi_f ) $ and 
$\chi_f / ( 1 - g_{bf}^2 \chi_b \chi_f )$
be positive.  Using the $\omega = 0$ values of
$\chi_b$ and $\chi_f$ above, these necessary conditions
can be rewritten as $g_{bb}> 0$ and $W \equiv N_f g_{bf}^2 / g_{bb} < 1$.
Using the expression of $N_f$ given earlier,  the last inequality gives
$ n_f ^{1/3} g_{bf}^2 < { 2 \over 3} A g_{bb}$ where
$A \equiv {\hbar^2 \over 2 m_f} ( 6 \pi^2 ) ^{2/3}$ as 
defined in \cite{Viverit00}.
These conditions were derived 
earlier in \cite{Viverit00} and \cite{Bijlsma00} using
slightly different considerations.   
For bosons and fermions with similar masses, 
we shall see shortly that $W$, a dimensionless parameter, 
serves as a useful measure
of the coupling between the bosons and fermions. 
If the bosons and fermions have similar masses,
$|W|$ is of order $ |a_{bf}^2 / a_{bb} n_f^{-1/3}| $ and thus typically
small for dilute gases unless $ |a_{bf}| >> |a_{bb}|$.
We shall limit ourselves only to the cases where $ |W|$'s are small.

We shall discuss now the behavior of $\tilde \chi_b$ and
$\tilde \chi_f$ in turn.  The results are qualitatively
different depending on whether $c_b {> \atop < } v_f$.
The velocity ratio $u \equiv c_b / v_f$
can be re-expressed as
$u = {m_f \over m_b} { (4/3)^{1/3} \over \pi^{1/6} }
{(n_b a_{bb})^{1/2} \over n_f^{1/3} } $.  
The value of $u$ can basically be arbitrary without
violating any stability criterion
(not only the linear stability condition above but
also others derived in \cite{Viverit00})

\noindent {\it Bosonic Response:}

\begin{figure}[h]
\epsfig{figure=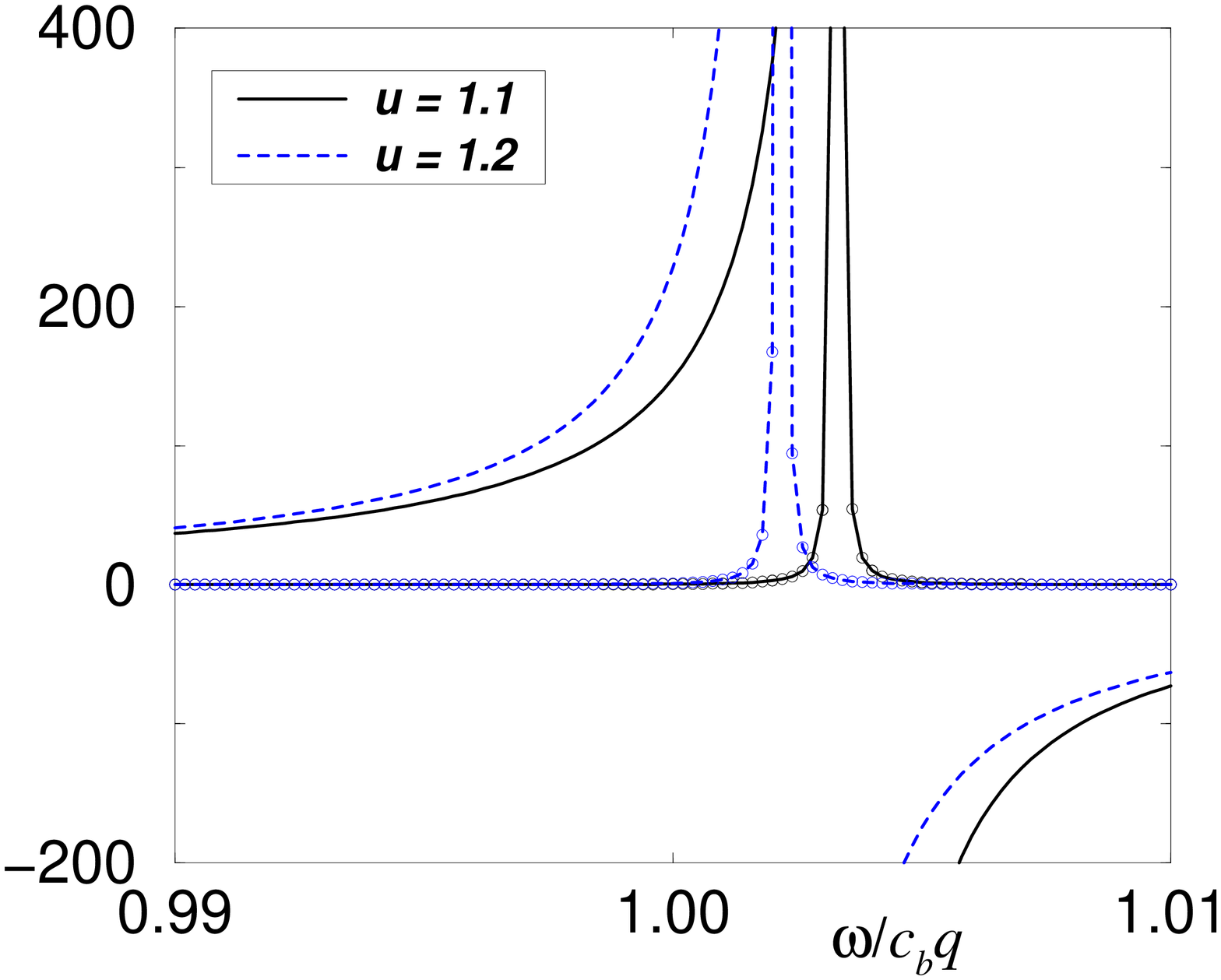,width=3in,angle=-90}
\vskip 0.4 cm
\begin{minipage}{0.45\textwidth}
\caption[]{ Dimensionless bosonic responses
${\rm Re} \tilde \chi_b$ and 
${\rm Im} \tilde \chi_b$ for $u \equiv c_b/v_f > 1$,
$ W = 0.01$.  Lines for the imaginary parts are decorated
with circles. }

\label{fig:buu}
\end{minipage}
\end{figure}

\noindent 1.  $c_b > v_f$:  In this regime a propagating
bosonic mode exists.  It can be easily verified ({\it e.g.}
graphically) that the mode frequency $\omega$ satisfies
$\omega > c_b q$ ( $> v_f q$).  The original bosonic
mode at $\omega = c_b q$ is pushed upwards by the
particle-hole `modes' lying below.  Some examples are
shown in Fig \ref{fig:buu}.  This mode `repulsion' is generally
expected ({\it c.f.} coupled harmonic oscillators).  It is
however of interest to examine the microscopic nature of the mode.
At the mode frequency both $\chi_b$ and $\chi_f$ are negative.
Thus, {\it e.g.}, if $g_{bf} > 0$, $\delta n_b$ and
$\delta n_f$ are of the same sign (see eqn (\ref{r2})).
The repulsion between the two species provides the
enhanced restoring force and oscillation frequency.
This frequency shift is typically small since usually $W <<1$.

\begin{figure}[h]
\epsfig{figure=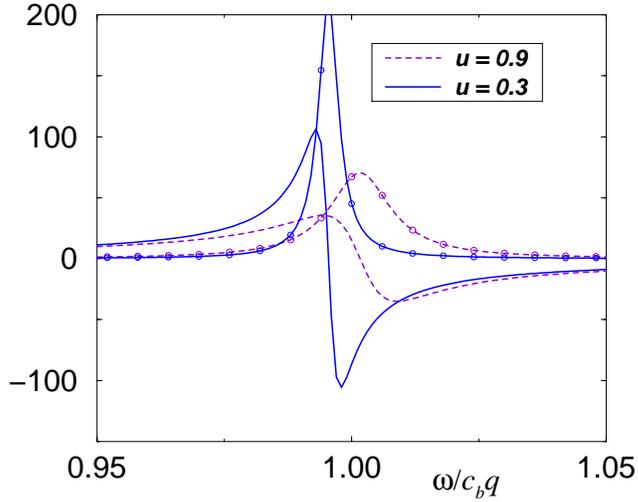,width=3in,angle=-90}
\vskip 0.4 cm
\begin{minipage}{0.45\textwidth}
\caption[]{ Same as Fig. 1 but for $u \equiv c_b/v_f < 1$, 
$W = 0.01$. }

\label{fig:bul}
\end{minipage}
\end{figure}

\noindent 2. $c_b < v_f$:  In this case the original bosonic
mode lies inside the particle-hole continuum of the fermions.
The bosonic mode is thus Landau damped.  For weak-coupling
the damping, thus the width of the response,
can be estimated easily using eq(\ref{bmode}) to
be $\sim \left[ {\pi N_f g_{bf}^2 \over 4 g_{bb} } \right] \ 
 \left[ { c_b \over v_f} \right] \ (c_b q) $.
Examples are shown in Fig \ref{fig:bul}.  
There is a small shift of the mode due to ${\rm Re} \chi_f$.
The shift is towards higher frequency for $u$ sufficiently
close to $1$ but opposite otherwise 
(${\rm Re \chi_f} \  < (>) \ 0 $ for 
${\omega / v_f q } >  (<) \ 0.83$.  )

\noindent 3.  It is also of interest to study the bosonic 
mode for $g_{bb} < 0$.  This is in fact the case for
the $^6$Li-$^7$Li mixture investigated in ref \cite{Mewes00},
 where the $^7$Li bosons have
a negative scattering length of $\approx - 1.5 {\rm nm}$.
In this case the original bosonic system is unstable,
and the Bogoliubov mode has an imaginary frequency
for sufficiently small wavevector ( $q < q_c = 2 m_b |c_b| / \hbar$,
here $ |c_b| \equiv [ |g_{bb}| n_b/m_b ]^{1/2}$).  
Since $N_f g_{bf}^2 > 0 > g_{bb}$, the system is still
unstable in the presence of the fermions 
\cite{Viverit00} (see also above).
Of interest is the effect of the fermions on the
unstable mode.  Now for imaginary frequencies $\omega = i \alpha$,
$\chi_f (q, i \alpha) = N_f \{ 1 - {\alpha \over v_f q} \
         [ {\pi \over 2} - {\rm tan}^{-1} {\alpha \over v_f q} ] \}$
is purely real and positive. $\chi_f$ decreases monotonically
with $\alpha$ from $\chi_f = N_f$ at $\alpha=0$ 
to $0$ as $\alpha \rightarrow \infty$.
 It can be easily verified that there is a {\it real}
solution for $\alpha$ to the dispersion relation ({\it c.f.} eq (\ref{bmode}) )

\be
[ \alpha ^2 - |c_b|^2 q^2 ]
- \left( {g_{bf}^2 \over |g_{bb}| }\right)
  ( |c_b|^2 q^2 ) \chi_f (q, i \alpha) = 0
\label{bimode}
\ee
for sufficiently small $q$  
(which includes the physically most relevant region
where $\alpha$ attains its maximum, i.e., the 
fastest growing instability).
Thus the instability is {\it not}
damped by the particle-hole degree of freedom.  In fact it
can be verified easily that, for given $q$, $\alpha$ is increased in
the presence of the fermions.  The system has become
even more unstable. This
mode has $\delta n_b$ and $\delta n_f$ of opposite signs
and corresponds to phase-separation as expected.

\noindent {\it Fermionic Response:}

\begin{figure}[h]
\epsfig{figure=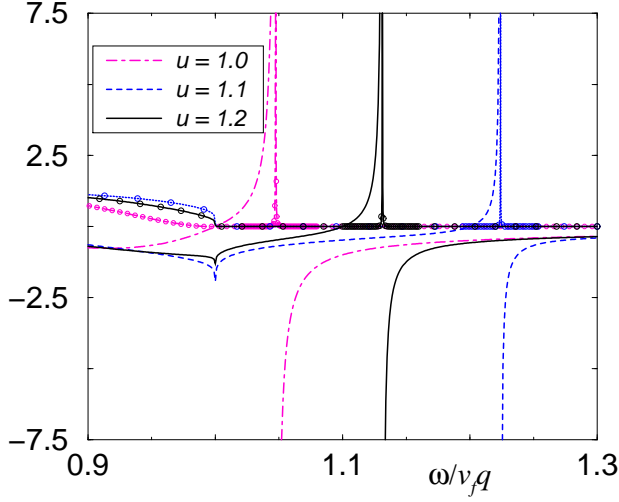,width=3in,angle=-90}
\vskip 0.4 cm
\begin{minipage}{0.45\textwidth}
\caption[]{ Dimensionless fermionic responses
${\rm Re} \tilde \chi_f$ and 
${\rm Im} \tilde \chi_f$ for $u \equiv c_b/v_f \ge 1$,
$W = 0.1$.
The imaginary parts (lines decorated with circles)
 contain the particle-hole continua
$\omega < v_f q$ and sharp spikes
at the bosonic mode frequencies. }

\label{fig:fu}
\end{minipage}
\end{figure}

\noindent 1. $c_b > v_f$:  In this case the fermionic response
for $0 < \omega < q v_f$ is only slightly modified.  A new 
feature appears near $\omega \sim c_b q > v_f q$ due to
the coupling to the bosonic mode.  An example is
as shown in Fig \ref{fig:fu}.  

\begin{figure}[h]
\epsfig{figure=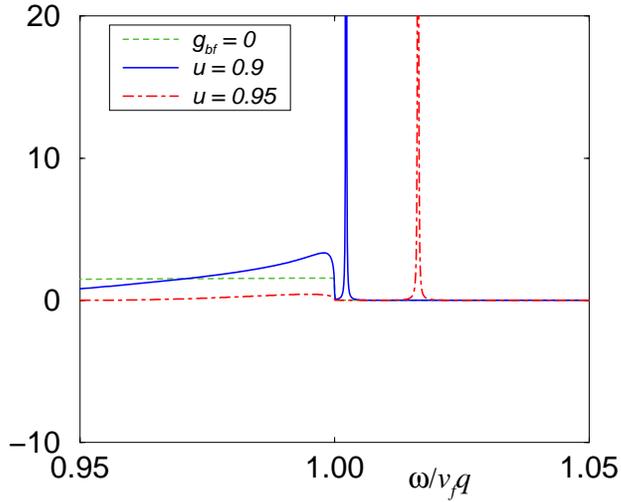,width=3in,angle=-90}
\vskip 0.4 cm
\begin{minipage}{0.45\textwidth}
\caption[]{ Dimensionless fermionic response  
${\rm Im} \tilde \chi_f$ for $u \equiv c_b/v_f < 1$
showing the zero-sound modes induced by the bosons.
$W = 0.1$.
Also shown is ${\rm Im} \chi_f$ for the pure
fermi gas ($g_{bf} = 0$) for comparison. }

\label{fig:imag}
\end{minipage}
\end{figure}

\noindent 2. $c_b < v_f$:  In this regime there are two
important features of the  fermionic response.  If 
$u = {c_b \over v_f}$ is sufficiently close to $1$,  the imaginary
part contains a sharp resonance at $\omega$ above the 
particle-hole continuum  (Fig \ref{fig:imag}).
There are two ways of understanding 
this mode.  It can be regarded as the continuation of
the situation from $c_b > v_f$, {\it i.e.}, it is due to
the bosonic mode which is itself slightly pushed up in frequency
({\it c.f.}, Fig \ref{fig:fu}, note in particular the result
for $u=1$).  Alternatively,
this mode can be considered as a zero-sound mode
induced by the bosons.  The form for $\tilde \chi_f$ in
eq (\ref{nf}) is precisely that of an {\it interacting} fermi gas
with s-wave interaction $g_{ff}$ (and therefore necessary
with more than one spin species,
 where the response is given by $\chi_f / ( 1 + g_{ff} \chi_f)$)
though with an effective {\it frequency dependent} interaction
$g_{ff} \rightarrow - g_{bf}^2 \chi_b$,  {\it i.e.} an effective
s-wave Landau parameter given by $F_0 \rightarrow
W / [\left( {\omega \over c_b q}\right)^2 - 1 ]$.
The bosonic mode $ \omega \sim c_b q$ for $c_b$ sufficiently close to but below
$v_f$ will thus induce a zero-sound mode for the fermions
just like an interaction among the fermions will. \cite{note}
 Note however
there cannot be a real s-wave interaction among the fermions
as they are of equal spins.  Thus this mode
{\it cannot} be obtained by considering the effective interaction 
among the fermions as in ref \cite{Bijlsma00}. 

The frequency of this propagating mode
can be estimated by using the well-known dispersion relation
of the zeroth sound 
$ \omega / v_f q \approx 1 + 2 e^{ -2 ( 1 + {1 \over F_0} ) } $
with the effective $F_0 \rightarrow  W / [ ({1 \over u})^2  - 1 ] $
as suggested above.  In order for the velocity of
the mode to be say $ 1 \%$ above $v_f$, then $c_b$
has to be within around $ 7 \%$ of $v_f$ if $ W = 0.1$.
This estimate agrees with the numerical results of Fig \ref{fig:imag}.  

\begin{figure}[h]
\epsfig{figure=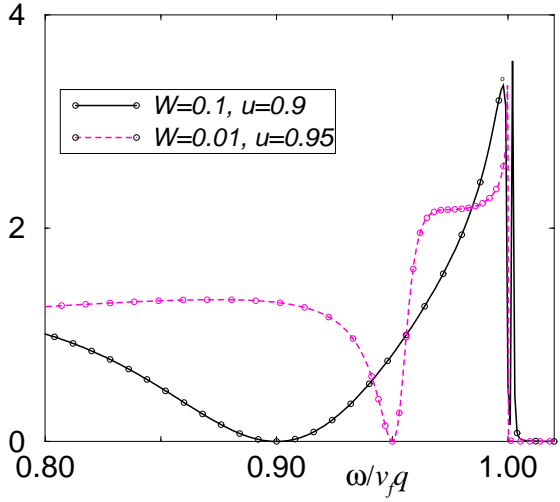,width=3in,angle=-90}
\vskip 0.4 cm
\begin{minipage}{0.45\textwidth}
\caption[]{ Imaginary part of dimensionless fermionic response  
${\rm Im} \tilde \chi_f$ for $u \equiv c_b/v_f < 1$,
showing mainly the region $\omega < v_f q$.  }

\label{fig:ful}
\end{minipage}
\end{figure}

The second interesting feature is that near the original 
bosonic mode frequency $\omega \sim c_b q$, there is
a reduction in the absorptive part $ {\rm Im} \tilde \chi_f$
(see Fig \ref{fig:ful}).
In fact ${\rm Im} \tilde \chi_f \rightarrow 0$ as $\omega \to c_b q$.
This, as well as the corresponding behavior of ${\rm Re} \tilde \chi_f$,
 can be seen easily mathematically from
eq(\ref{r2}) due to the resonance nature of $\chi_b$ at this frequency.
Physically this can be regarded as due to mode-mode repulsion --
the bosonic mode has pushed away the particle-hole `modes' near
$\omega \sim c_b q$.  This feature is present even
for small coupling $W$.  A larger $W$ mainly increases the 
width of this `transparent' region.  Thus in fact
the frequency dependence of ${\rm Re} \tilde \chi_f$ is
actually {\it stronger} for smaller $W$'s.

The energy absorption by the bose-fermi mixture 
 from an external perturbation acting on the fermions 
is thus substantially
reduced for frequencies within this 
 `transparent region'. The width of this region can be estimated
by using the observation that the fermionic response
is roughly reduced by the factor $ 1 + W { (c_b q)^2 \over
  \omega^2 - (c_b q )^2 } $  for these frequencies.
For the fermionic response to be reduced to say $1/2$ of
its bare value, then  $ | \omega - c_b q | / c_b q < W /2$.
This estimate agrees very roughly 
with the numerical results in Fig. \ref{fig:ful}.

In conclusion,  I have investigated the collective modes
of a Bose-Fermi mixture, and have shown that there
is important mode-mode coupling effects, especially if
$v_f \sim c_b$.

I thank David Edwards for helpful correspondences.
This research was supported by the National Science
Council of Taiwan under grant number 89-2112-M-001-105.

\begin{figure}[h]
\epsfig{figure=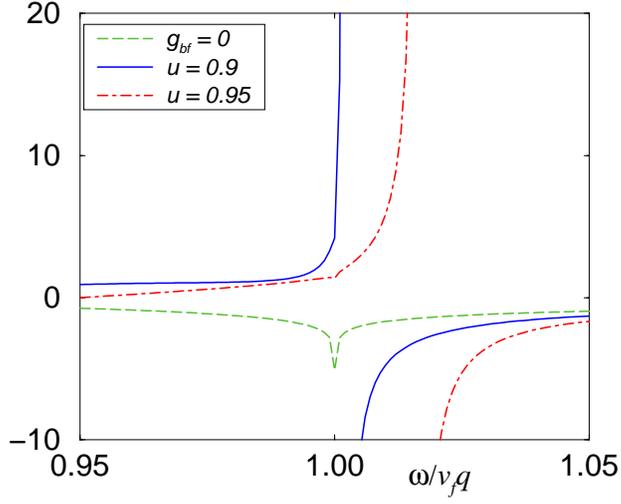,width=3in,angle=-90}
\vskip 0.4 cm
\begin{minipage}{0.45\textwidth}
\caption[]{ Same as Fig \ref{fig:imag} except
that now 
${\rm Re} \tilde \chi_f$ is shown. }

\label{fig:real}
\end{minipage}
\end{figure}

\end{document}